\documentclass[12pt]{irsreport}
\usepackage{epsfig}

\title{IRS-TR 12002:  Constructing a Short-Low Truth Spectrum 
of the Standard Star HR 6348}

\author{
G.C. Sloan (1) ~\& D. Ludovici (2) \thanks{ (1) Infrared Spectrograph 
Science Center, Cornell University, (2) Department of Physics and
Astronomy, University of Iowa; NSF REU Research Assistant, Astronomy 
Department, Cornell University} }

\date{14 December, 2012}

\runninghead{IRS-TR 12002 - The SL Spectrum of HR 6348}

\begin{document}

\maketitle

\begin{abstract}

This report describes in detail the generation of a ``truth''
spectrum of HR~6348, using observations with the Short-Low 
(SL) module of the Infrared Spectrograph of HR~6348, and the 
A dwarfs $\alpha$~Lac and $\delta$~UMi.  Using spectral
ratios, we can propagate Kurucz models of the A dwarfs to the 
K giant HR~6348, which can then serve to calibrate the 
remaining database of SL spectra.  Mitigation in the vicinity 
of the Pfund-$\alpha$ line is necessary to reduce residual 
artifacts at 7.45~\mum.  In general, the new SL spectrum
of HR~6348 has a spectroscopic fidelity of $\sim$0.5~\% or 
better.  Artifacts from the hydrogen recombination lines in
the A dwarfs will generally be smaller than this limit, 
although the residual artifact from the blend of lines near 
Pfund-$\alpha$ exceeds the limit at $\sim$0.7~\%.

\end{abstract}

\section{Introduction} 

The Infrared Spectrograph (IRS; Houck et al.\ 2004) on the 
{\it Spitzer Space Telescope} (Werner et al.\ 2004) operated
from launch in August, 2003 to the exhaustion of onboard
cryogens in May, 2009.  During this period, the IRS obtained
over 16,000 low-resolution spectra, most with both the
Short-Low (SL) and Long-Low (LL) modules.  The cryogenic
mission included hundreds of observations of standard stars,
most of which were A dwarfs or K giants.  The use of standard 
stars to calibrate the spectra of other sources uses the 
equation
\begin{equation}
S_T = \frac{S_O}{C_O} \, C_T ,
\end{equation}
where the subscript ``T'' denotes the actual spectrum of a 
target, the subscript ``O'' denotes its observed spectrum, 
``C'' refers to the calibrator (or standard star), and ``S'' 
refers to the science target (the source being calibrated).
The spectral correction needed to convert raw spectra from 
the SL module into accurate flux-density units is often
referred to as the relative spectral response function, or 
RSRF.  It is the ratio $C_T/C_O$.  Any errors in the assumed 
spectrum $C_T$ for a chosen standard will propagate to the 
entire spectral database.  The spectrum $C_T$ is commonly 
referred to as a ``truth'' spectrum, although it will never
be perfect.  Our objective is to minimize the imperfections 
in the truth spectra for our chosen standards, because those 
imperfections will propagate to the full SL database as 
artifacts.

The IRS Team at Cornell chose HR~6348 as the primary standard 
for SL and LL for several reasons.  First, it is relatively 
bright, allowing quick integrations to obtain high 
signal/noise (SN) ratios.  Second, it is faint enough to use 
the IRS Red Peak-Up sub-array for target acquisition, which 
improves our ability to center it in the spectroscopic slit 
and provides contemporaneous infrared photometry.  Third, it 
is close enough to the northern continuous viewing zone (CVZ) 
to allow observations in nearly all IRS observing campaigns.
And fourth, it is an early K giant with relatively weak
molecular bands, most notably from SiO at 8~\mum.  Later
K giants have stronger bands which make their use as 
standards more challenging.

Prior to launch, the IRS Teams at Cornell and the {\it 
Spitzer} Science Center (SSC) decided to rely on two methods 
to estimate truth spectra:  (1) synthetic spectra 
generated from stellar models, and (2) spectral templates, 
which are based on observed truth spectra of stars with 
identical spectral classifications.  As described by Sloan et 
al.\ (2013), neither method proved suitable.  Synthetic 
spectra do not predict the depth of molecular bands in 
late-type stars with sufficient accuracy, although they work 
well for early-type stars with only atomic lines in their 
spectra.  Additionally, the variations in the depths of 
molecular bands within a given spectral class are large 
enough to make spectral templates too inaccurate.

This report describes in detail the generation of a fully
calibrated spectrum of HR~6348, to be used as the assumed
spectrum when using observations of HR~6348 to determine 
the spectral correction (or RSRF) for SL.  Without a means
to estimate its truth spectrum apriori, we chose to repeat
the methodology of Cohen et al.\ (1992a,b,c), who calibrated 
the spectrum of the K5 giant $\alpha$~Tau using similar
observations and models of the A dwarfs $\alpha$~CMa and
$\alpha$~Lyr.  Over the next decade, they built a library
of 16 calibrated spectra of standard stars, any of which 
could serve as a ``truth'' spectrum (Cohen et al.\ 1995, 
1996a, 1996b, 2003).

The general plan for HR~6348 follows theirs.  Using Kurucz 
models of the A dwarfs $\alpha$~Lac and and $\delta$~UMi, we 
first calibrate the low-resolution IRS spectrum of HR~6348.
Then using this standard star, we can calibrate additional
standards and propagate the calibration to the entire 
database of low-resolution IRS spectra.  This report
concentrates on the spectra in the SL module, which covers
the 5--14~\mum\ range.  IRS-TR 12003 (Sloan \& Ludovici 
2012b) will describe the calibration of HR~6348 using data 
from LL and the combination of data from SL, LL and the 
Red-PU sub-array to produce a photometrically calibrated 
5--37~\mum\ spectrum of HR~6348.

\section{Kurucz models} 

We begin with a Kurucz model of $\alpha$~Lac supplied to the
IRS Team by M.\ Cohen.  The team received the original 
low-resolution model prior to launch; they received a 
high-resolution model with $R$ ($\lambda$/$\Delta \lambda$) 
= 1000 in September, 2004.  To shift this model to the
wavelength grid and resolution of SL and LL, we convolved it 
with gaussians with the widths given by Table~1.  These 
gaussians were chosen to reproduce the widths of the hydrogen 
recombination lines as observed in SL and LL.

\bigskip
\begin{center}
\begin{tabular}{lc} 
\multicolumn{2}{c}{\bf Table 1---Convolution Parameters} \\
\hline
{\bf Order} & {\bf Gaussian width (in pixels)} \\
\hline
SL2              & 0.75 \\
SL1 and SL-bonus & 1.00 \\
LL2              & 0.95 \\
LL1 and LL-bonus & 0.75 \\
\hline
\end{tabular}
\end{center}
\bigskip

We also shifted the models photometrically to match the
Red PU photometry as calibrated by IRS-TR 11002 (Sloan \&
Ludovici 2011).  This latter step is not essential, because 
the loss of flux in some pointings in SL will require that we
rescale the final SL spectrum of HR~6348 to match LL.

In order to fit the available optical and infrared 
photometry, the Cohen-supplied model is based on an A2 dwarf,
even though the spectral class of $\alpha$~Tau is A1~V.  The
interstellar extinction ($A_v$) is zero.  The low-resolution
model for $\delta$~UMi (A1~Vn) differs from $\alpha$~Lac
only in its photometric scaling.  Because we did not have a
high-resolution model for $\delta$~UMi, we used the model for
$\alpha$~Lac, scaled to match the Red PU photometry.

\section{Procedure} 

To calibrate the spectrum of HR~6348, we followed the same
procedure in all three of the available orders.  We 
calibrated HR~6348 in each of the two nod positions in each 
order.   For each nod position, we calibrated the data 
twice, once with $\alpha$~Lac and once with $\delta$~UMi.
Within each nod, we combined the separately calibrated data
by forcing them to a spline-smoothed average while preserving 
their separate noise (and artifact) characteristics.  When 
combining the two spectra to a single nod spectrum, we set 
the weighting for one of the spectra to zero if it showed 
more structure than the other.  We followed an identical 
technique when combining the two nod spectra into one 
spectrum for the order in question.  Generally, this approach
conserves the overall shape of the spectrum, while at each
stage, it selects for the smoother input.  The next three 
sections describe the procedure for SL1, SL2, and the 
SL-bonus order (SL-b), respectively.

We did not use all of the available data for any of the three
stars considered.  IRS-TR 12001 (Sloan \& Ludovici 2012a) 
describes how the level and overall shape of a spectrum can 
vary depending on its position within the slit.  These 
spectral pointing-induced throughput errors, or SPITE, have 
been a concern since before {\it Spitzer} launched.  We 
reject any pointings if either the total throughput loss in 
SL1 is 8\% of more, or if the change in color, as measured 
between the 7.5--9.5~\mum\ and 11.5--14.0~\mum\ windows, is 
more than 1\%.  Additional images might also face rejection 
for a variety of reasons, including data drop-outs and 
residual images from previously observed bright sources.

In SL, the only suspicious artifact on the order of $\sim$1\%
or larger appears at 7.45~\mum, at the position of a blend
of the Pfund-$\alpha$ line with other hydrogen recombination
lines.  Sec.~7 explains how we mitigated for this artifact.

Our spectrum is only valid in the wavelength regimes for
which the data are valid.  However, it is advantageous to
make a reasonable estimate of the spectrum at all wavelengths
where SL obtained data, even if of dubious quality.  Sec.~7 
also describes how we filled out the available wavelength 
space.

\section{Short-Low Order 1} 

\begin{figure} 
  \begin{center}
     \epsfig{file=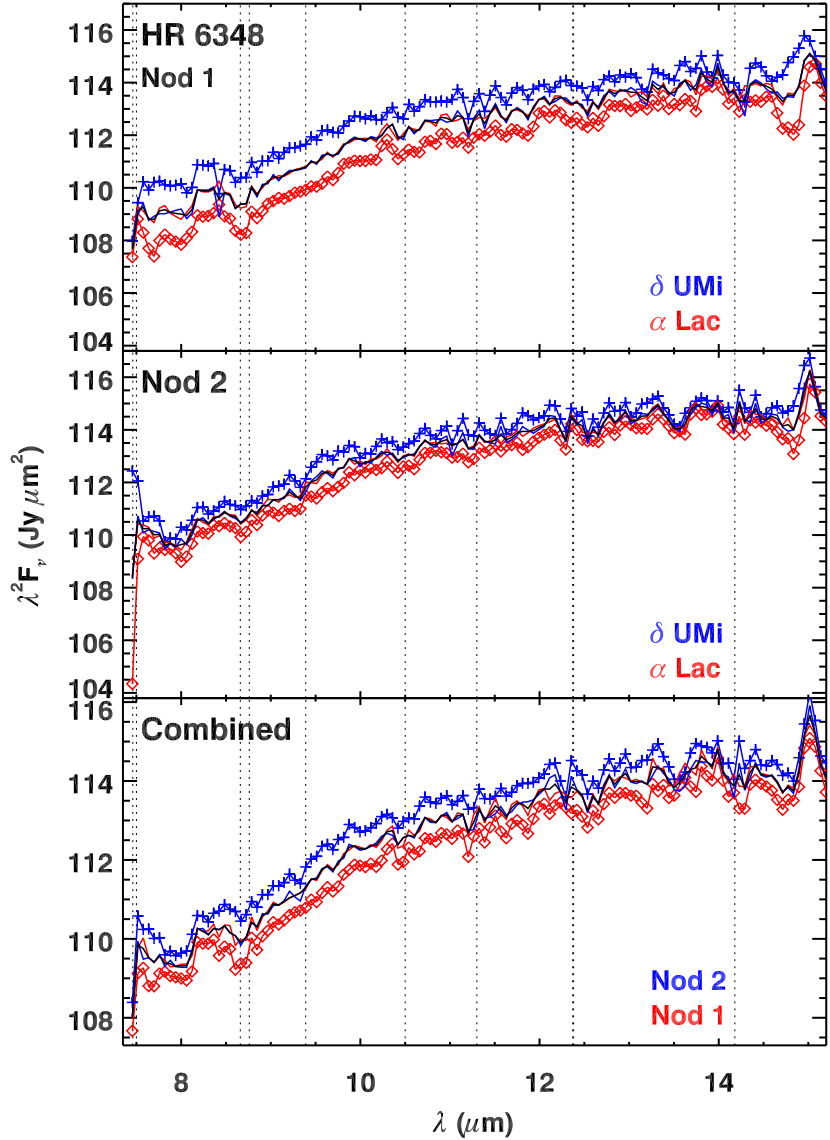, width=11cm}
  \end{center}
\caption{
---The construction of the SL1 spectrum of HR~6348.  In each
panel, the colored traces with symbols show the input
spectra, while the colored traces without symbols show the
spectra after they have been normalized to the 
spline-smoothed average.  The black trace shows the 
weighted average of the spline-normalized spectra.  In the
top two panels, the input spectra are color coded depending 
on whether they were calibrated with $\alpha$~Lac or 
$\delta$~UMi.  In the bottom panel, the input spectra are
from the two nods.  The vertical dotted lines mark the 
positions of the strongest hydrogen recombination lines in 
the spectrum.  From left to right, these are:  
Pfund~$\alpha$ (7.46~\mum), 8--6 and 11--7 (7.50~\mum), 
14--8 (8.66~\mum), 10--7 (8.76~\mum), 13--8 (9.39~\mum), 
12--8 (10.50~\mum), 9--7 (11.30~\mum), 7--6 and 11--8 
(12.37--12.38~\mum), and 13--9 (14.18~\mum).}
\end{figure}

Figure~1 illustrates the construction of the SL1 spectrum of
HR~6348.  The top two panels show each nod separately.  
Generally, when one calibration showed more structure than 
the other in a given wavelength range, we used only data for 
the other.  For Nod 1, we ignored the $\alpha$~Lac-calibrated 
spectrum at 8.60--8.73 and 10.24-10.36~\mum.  For Nod 2, we 
dropped the $\alpha$~Lac-calibrated spectrum at 
8.91--9.09~\mum\ and the $\delta$~UMi-calibrated spectrum at 
11.02--11.14~\mum.  When combining the nods, we dropped Nod 1
at 8.30--8.60, 8.73--8.85, and 10.42--10.54~\mum\ and Nod 2 
at 8.18--8.24 and 9.21-9.33~\mum.

\section{Short-Low Order 2} 

\begin{figure} 
  \begin{center}
     \epsfig{file=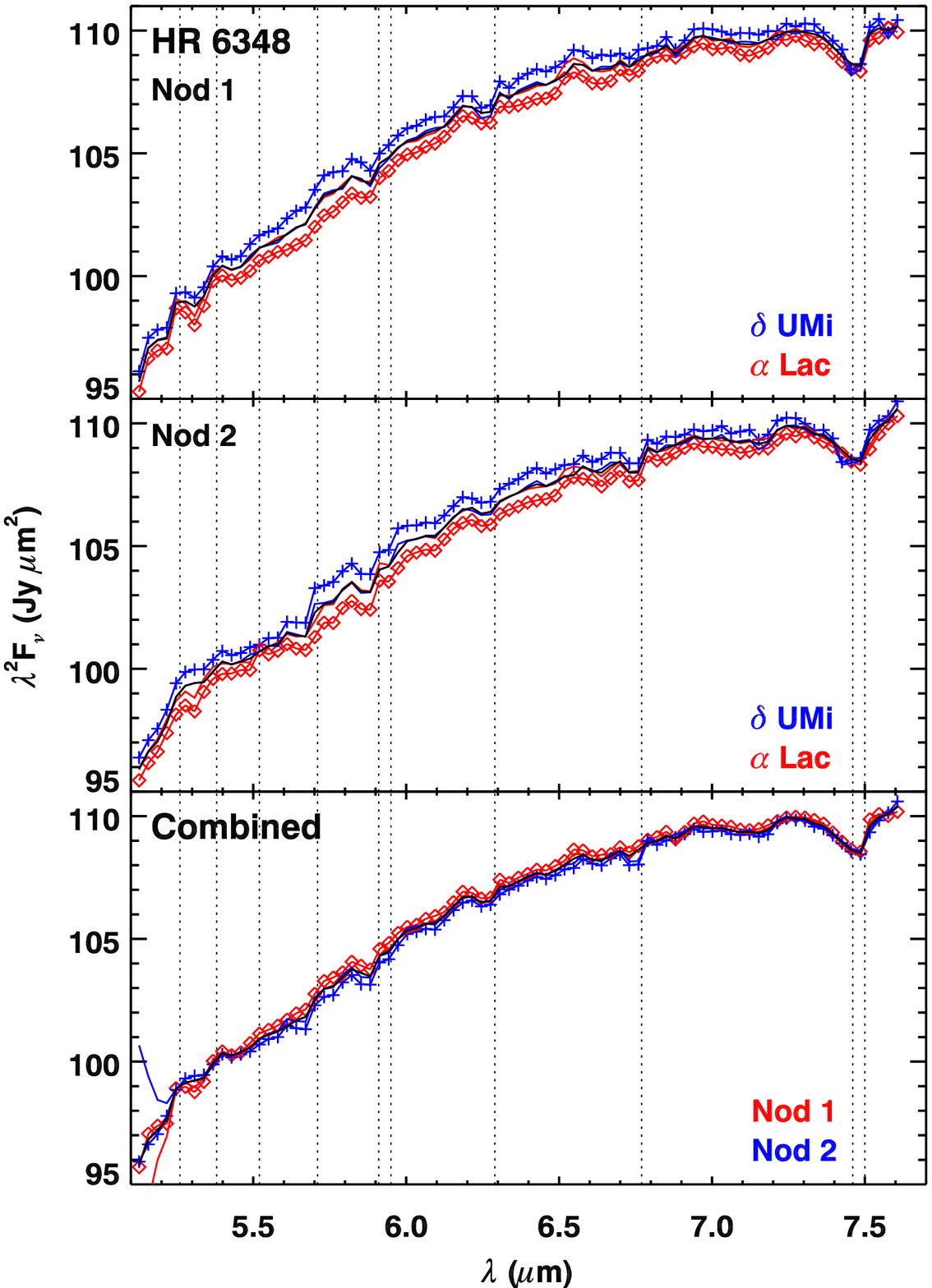, width=11cm}
  \end{center}
\caption{
---The contruction of the SL2 spectrum of HR~6348.  The
panels are as described in Fig.~1.  The vertical dotted lines 
show the following hydrogen recombination lines:  18--7 
(5.26~\mum), 17--7 (5.38~\mum), 16--7 (5.52~\mum), 15--7 
(5.71~\mum), 9--6 (5.91~\mum), 14--7 (5.95~\mum), 13--7 
(6.29~\mum), 12--7 (6.77~\mum), Pfund~$\alpha$ (7.46~\mum), 
and 8--6 and 11--7 (7.50~\mum).}
\end{figure}

Figure~2 illustrates the construction of HR~6348 in SL2, 
following the same procedure as in SL1.  Because the first
four pixels (5.13--5.22~\mum) of the $\alpha$~Lac calibration
created problems for the spline-fitting algorithm, we replaced 
these with the corresponding data from the $\delta$~UMi
calibration (keeping the nods separate).  As before, data in 
a pair being combined showing more structure than its
counterpart were ignored.  When combining the spectra for Nod 
1, we discounted the $\alpha$~Lac calibration at 5.25--5.43 
and 6.52--6.58~\mum\ and the $\delta$~UMi calibration at
5.91--5.97, 6.22--6.28, and 7.42--7.49~\mum.  For Nod 2, the
rejected data included the $\alpha$~Lac calibration at 
5.19--5.43, 5.52, 5.91-5.94, and 6.52--6.58~\mum\ and 
$\delta$~UMi calibration at 5.97--6.00, 6.22--6.28, and 
7.42--7.49~\mum.  When combining the nods, we rejected the
Nod 1 data at 5.31--5.34 and 5.43--5.49~\mum\ and the Nod 2
data at 5.61--5.67, 5.82--5.88, and 6.73--6.79~\mum.

\section{Short-Low Bonus Order} 

\begin{figure} 
  \begin{center}
     \epsfig{file=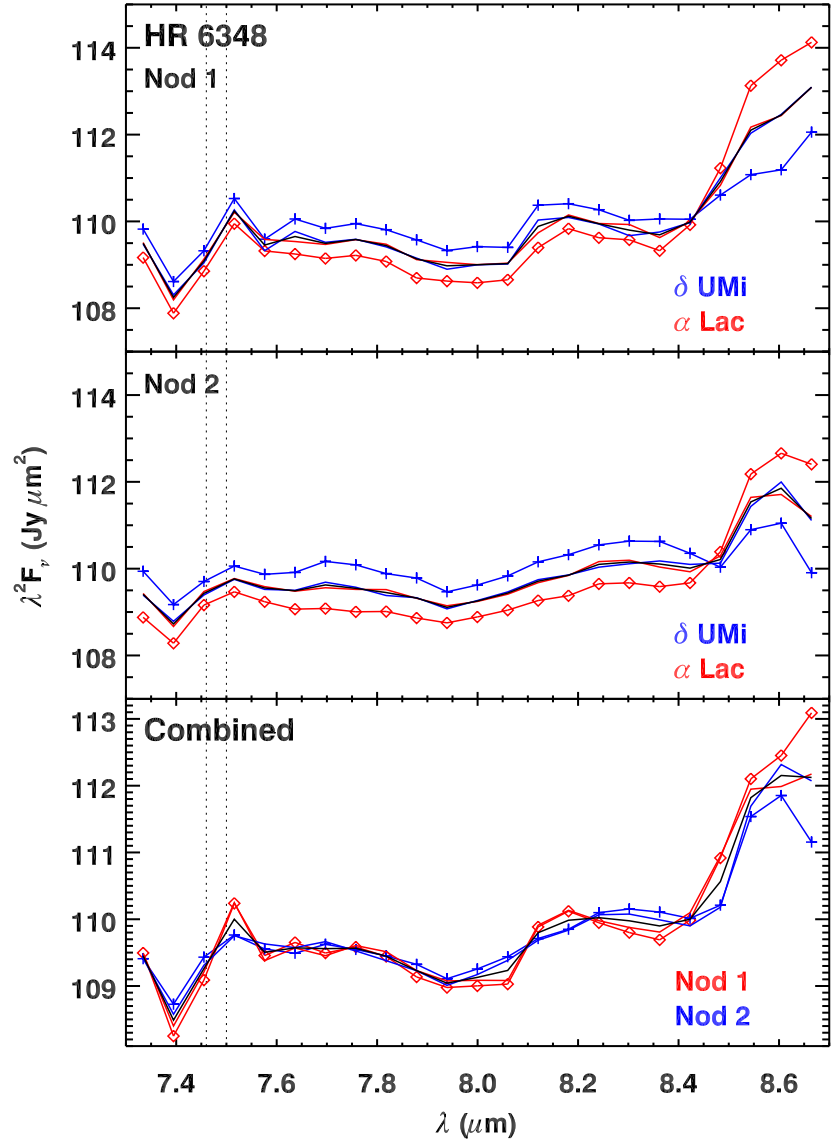, width=11cm}
  \end{center}
\caption{
---The contruction of the SL-bonus spectrum of HR~6348.  The
panels are as described in Fig.~1.  The vertical dotted lines 
depict the Pfund-$\alpha$ line (7.46~\mum), and 8--6 and 
11--7 recombination lines (7.50~\mum).}
\end{figure}

Figure~3 presents the construction of HR~6348 in the 
SL-bonus order.  Unlike the other two orders, none of the
data were rejected.  We simply used everything when combining
calibrations to make the two nods, and when combining the two
nods to make a single spectrum.

\section{Repairing Artifacts and Segment Ends} 

\begin{figure} 
  \begin{center}
     \epsfig{file=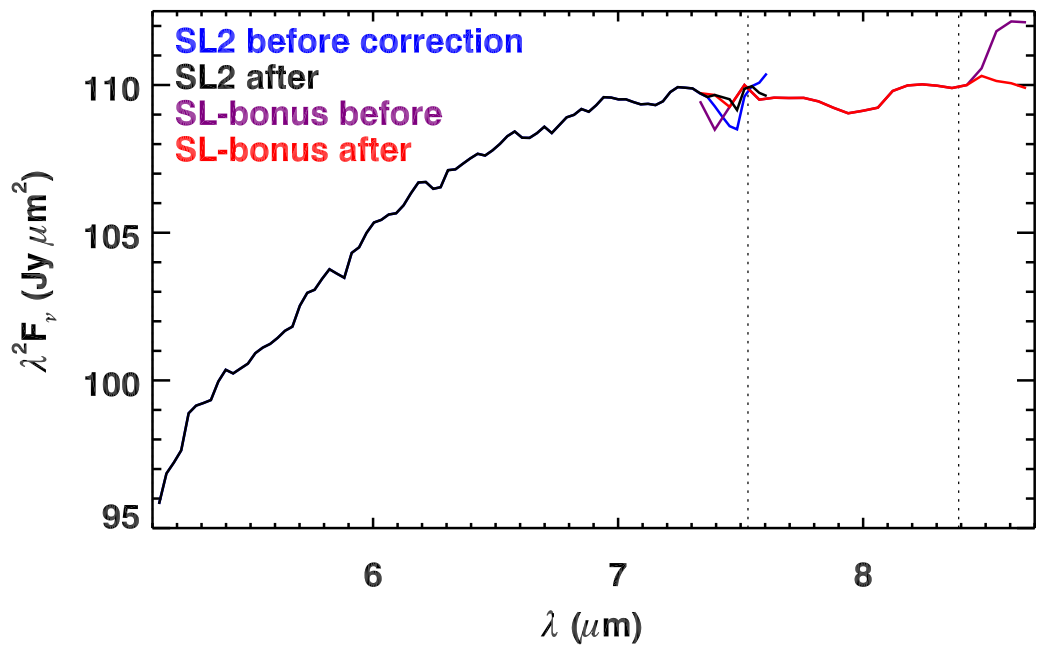, width=13.8cm}
  \end{center}
\caption{
---Correcting the Pfund-$\alpha$ artifact at 7.5~\mum\
and filling out the ends of SL2 and the SL-bonus order.
The vertical dotted lines mark the boundary between
useful data for SL2 and SL-b at 7.53~\mum\ and the red 
edge of the useful data for SL-b at 8.39~\mum.  The
SL2 data at 7.5~\mum\ (in blue) are corrected using data 
from red standards.  The bonus-order data outside the
range 7.53--8.39~\mum\ are replaced with either SL2
(corrected for Pfund-$\alpha$) or SL1.  The SL2 data
to the red of 7.53~\mum\ are replaced with SL1 data.}
\end{figure}

\bigskip
\begin{center}
\begin{tabular}{lcc} 
\multicolumn{2}{c}{\bf Table 2---Wavelength Ranges} \\
\hline
{\bf Spectral} & {\bf Full range} & {\bf Useful range} \\
{\bf segment} & {\bf \mum} & {\bf \mum} \\
\hline
SL2      & 5.13--7.61  & 5.13--7.52 \\
SL-bonus & 7.33--8.66  & 7.76--8.36 \\
SL1      & 7.46--15.26 & 7.58--14.17 \\
\hline
\end{tabular}
\end{center}
\bigskip

In Figures~1--3, the vertical dotted lines mark the 
positions of the more significant hydrogen recombination 
lines in the A dwarfs.  Much of the spectral structure in our
calibration of HR~6348 is in the vicinity of these lines,
suggesting that our low-resolution truth spectrum for the
A dwarfs is imperfect.  The possible artifacts are more 
noticeable in SL2, especially at 7.5~\mum, where the 
Pfund-$\alpha$ line is blended with the 8--6 and other 
transitions.  We estimate that the Pfund-$\alpha$ artifact 
has a peak strength of $\sim$1.5\%.

Most of the possible artifacts have a strength of $\sim$0.5\% 
or less, and because they are generally offset to the blue 
with respect to the expected line centers, we have chosen to
take the conservative approach and leave them alone.  This
decision sets a limit of $\sim$200 for a meaningful 
signal/noise ratio for SL data.

The artifact at 7.5~\mum\ clearly exceeds our 0.5~\% limit
and must be dealt with.  With an artifact this pronounced, 
we can generate a correction from other data.  We selected
the asteroids Amalia and Isara and the ultraluminous infrared 
galaxy IRAS 07598+6508 as our test set.  All three sources 
were observed by the IRS as red calibrators, and none show
any spectral structure in the vicinity of the Pfund-$\alpha$
line.  We calibrated each source using the calibrated 
spectrum of HR~6348, as it appears in the bottom panels of 
Figures~1--3, as the truth spectrum.  All three show the 
Pfund-$\alpha$ artifact, and from these we constructed a 
correction running from 7.33 to 7.52~\mum, peaking at 0.8~\% 
at 7.46~\mum.  This correction removes over half of the 
Pfund-$\alpha$ artifact in the spectrum of HR~6348.  

We are unable to remove any more of the artifact without
making a priori assumptions about what the spectrum of 
HR~6348 should look like in this region.  This region is
complicated not just by Pfund-$\alpha$, but also by the
fact that it coincides with the boundary between SL2 and
SL1, and because it marks the blue edge of the SiO 
absorption bandhead commonly seen in K giants.

It is useful to have a reasonable estimate for the truth
spectrum at all wavelengths covered by SL, even if the
module does not return meaningful data at those wavelengths.
Where spectral segments overlap their neighbors, we simply
use the neighboring data to fill them out.  We extrapolated
SL1 past its long-wavelength limit of 14.20~\mum\ using a
4400 K Engelke function (Engelke 1992).  These steps will
prevent data in these regions from being wildly 
miscalibrated, which is useful even if they ultimately will
be discarded.

\section{Quality Assessment} 

\begin{figure} 
  \begin{center}
     \epsfig{file=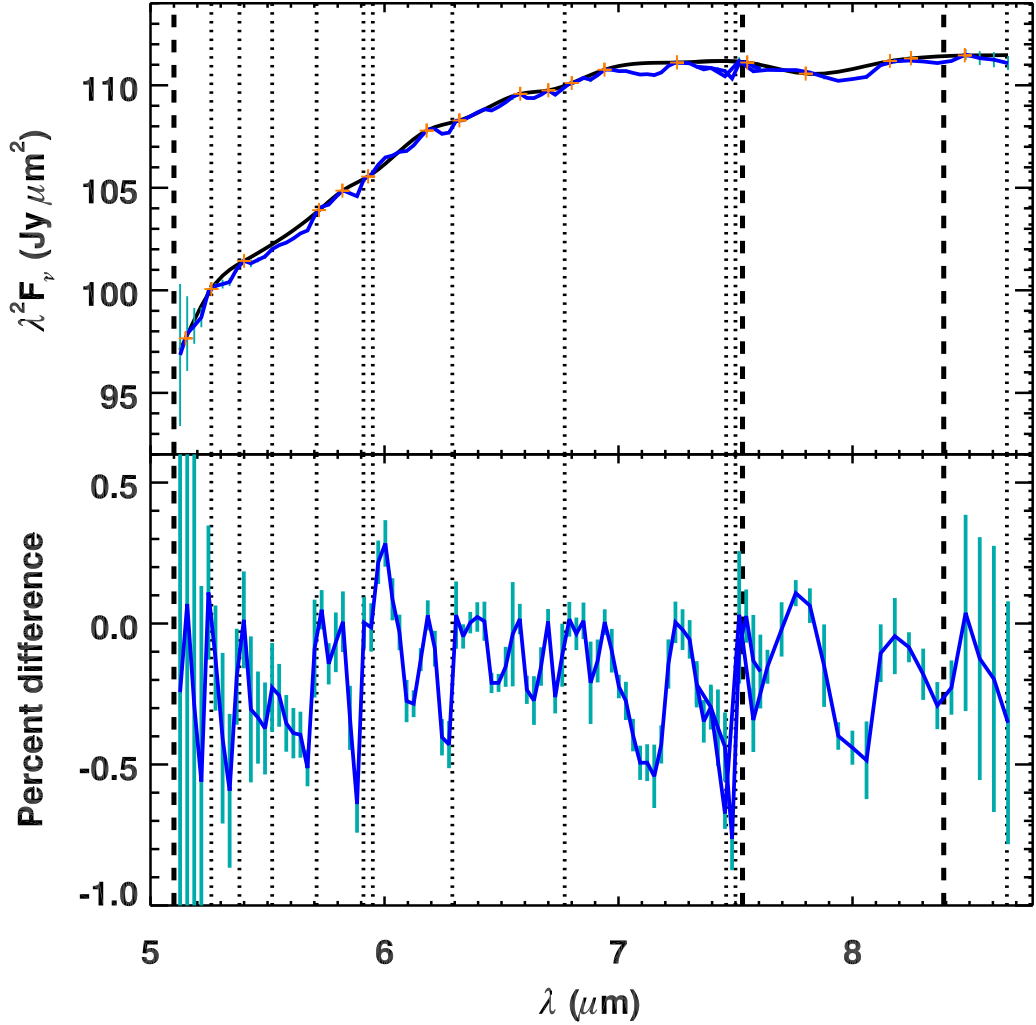, width=13.8cm}
  \end{center}
\caption{
---Estimating the strength of residual structure in the SL2
and SL-bonus spectra of HR~6348.  The top panel shows the 
actual spectrum (in blue, with uncertainties in light blue), 
the fitted spline (in black), and the spline points (as 
orange plus signs).  The bottom panel shows the difference
between the actual spectrum and the fitted spline, as a
percentage of the spline.  The thick dashed lines mark the
boundaries between spectral segments, and the thin dotted
lines mark the positions of recombination lines.  The
absorption at the position of the Pfund-$\alpha$ line has a
strength of $\sim$0.7\%.  The remaining features are 
generally smaller than $\sim$0.5\%.}
\end{figure}

\begin{figure} 
  \begin{center}
     \epsfig{file=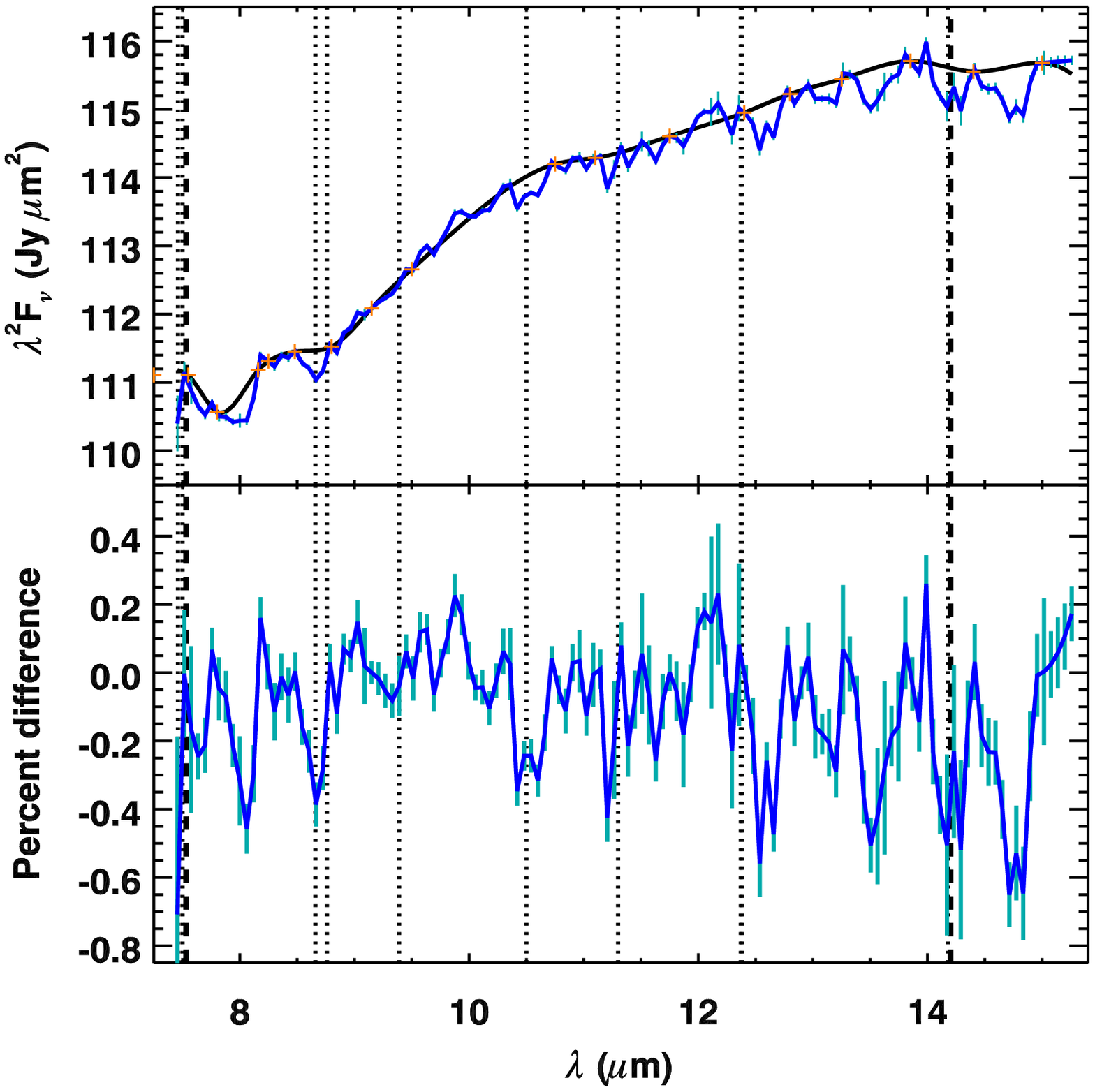, width=13.8cm}
  \end{center}
\caption{
---Estimating the strength of residual structure in the SL1
spectrum of HR~6348.  The panels, plotting symbols, and 
colors are as defined in Figure 5.  The absorption features
at 12.5, 13.0, 13.5, and 14.2~\mum\ are consistent with the
positions of OH absorption bands.  We fit the spline right
through the SiO absorption band at 8~\mum, leaving apparent
absorption features at 8.0 and 8.6~\mum.  It is likely that
all of the structure between 7.5 and 9.0~\mum\ in the upper
panel is real.}
\end{figure}

Figures~5 and 6 use fitted splines to estimate the strength
of the residual structure in our spectrum of HR~6348.  In
SL2, the most significant feature is at the position of the
Pfund~$\alpha$ line at 7.5~\mum.  Given the proximity of
this feature to the SL2/SL1 interface, we cannot state
whether the structure in our spectrum is an artifact or
really intrinsic to HR~6348.  Its strength is $\sim$0.7\%.
There appear to be artifacts in the vicinity of the
hydrogen recombination lines at 5.9 and 6.3~\mum, which may
arise from differences between the actual A dwarf spectra 
and our model.  These differences could arise from 
differences in line strengths, but it is also possible that
differences in line profiles from our assumed gaussians could 
be responsible.  

Some of the absorption features in HR~6348 do not coincide
with hydrogen recombination lines, such as the feature at
7.1~\mum\ and the smaller features at 6.1, 6.5, and 6.6~\mum. 
Whether these features are due to real structure in HR~6348
or to imperfections in the models of A dwarfs is an open
question.  They are within our stated envelope of $\sim$0.5\%
fidelity in our spectral calibration.

Figure~6 focuses on SL1.  We chose to force the spline 
through much of the weak SiO band in HR~6348, but one should
keep in mind that most of the structure between 7.5 and 
9.0~\mum\ is likely to be real and due to SiO.  The 
absorption feature at 8.6~\mum\ is close to the 14--8 and 
10--7 hydrogen transitions and is more debatable, but if it
is an artifact, its strength is only 0.3\%.  The broad 
absorption features at 12.5, 13.0, 13.5, and 14.2~\mum\ 
probably arise from OH bands and are quite likely real.  
The four most dominant absorption features between 9 and 
12~\mum\ are all close to the position of hydrogen 
recombination lines and may well be artifacts.  Their
strength is generally only $\sim$0.3\%.

We conclude that the spectral fidelity of the SL spectrum of 
HR~6348 presented here is generally better than $\sim$0.5\%
in SL2 and $\sim$0.3\% in SL1.  The notable exception is in
the vicinity of the Pfund-$\alpha$ line at 7.5~\mum, at the
junction between SL2 and SL1, where the fidelity is only
0.7\%.

\end{document}